\documentclass[12pt, a4paper]{article}

\begin{document}
\title{Interacting many-body systems as
non-cooperative games}
\author{Chiu Fan Lee\thanks{c.lee1@physics.ox.ac.uk}
\ and \ Neil F. Johnson\thanks{n.johnson@physics.ox.ac.uk}
\\
\\ Centre for Quantum Computation and Physics Department \\ Clarendon
Laboratory, Oxford University \\ Parks Road, Oxford OX1 3PU, U.K.}

\maketitle

\abstract{ We explore the possibility that physical phenomena
arising from interacting multi-particle systems, can be usefully interpreted
in terms of multi-player games. We show how non-cooperative
phenomena can emerge from Ising Hamiltonians, even though the individual
spins behave cooperatively. Our findings establish a mapping
between two fundamental models from condensed matter physics
and game theory.}

\newpage

The spatial prisoner's dilemma game of Nowak and May \cite{NM92}
showed that complex macroscopic patterns can develop and evolve in a system
even when the microscopic interactions are local in space and
time \cite{NM92,SH02,ST98}. This is reminiscent of the
observation in multi-particle interacting systems in condensed matter
physics (e.g. a spatial array of spins) whereby different macroscopic
`phases' can emerge. In a separate development, multi-spin models have been
shown to offer new tools and viewpoints to help in understanding problems
in information processing and optimization problems \cite{nish}.
Meanwhile multi-agent games have recently caught physicists' attention
via Econophysics
\cite{econ}; just as in more conventional condensed matter systems, it has
been shown that a renormalization of inter-agent interactions to form
non-interacting `crowd-anticrowd' clusters provides a good description
of the macroscopic fluctuations
\cite{michael}. 
Game-theoretic language has also emerged in the quantum
domain via the new field of quantum games \cite{qgames}.
Of relevance to
biologically-motivated physics is the finding that
mesoscopic biological systems such as viruses can play games, in particular
the fundamental  prisoner's dilemma game \cite{TC99}. Furthermore, it is
known that such games can exhibit a
variety of non-linear dynamical phenomena when evolved in time \cite{We95}.

Given all this cirumstancial evidence, is there possibly a deeper connection
between games and physical phenomena? At first sight, games and physics do
appear to have some common elements: both consider collections of
interacting objects following certain local rules which can give rise to a
number of macroscopic configurations. However there is a fundamental problem
with the analogy. Physical systems are always driven to minimize the free
energy hence the global optimal solution is preferred: particles'
behavior is therefore cooperative. On the other hand, the games which are
arguably most interesting and important are non-cooperative: as we explain
below, this means that their Nash equilibria are not Pareto optimal.

This Letter shows that non-cooperative behavior can indeed emerge in
physical systems, even with the individual particles behaving
cooperatively. We specifically consider Ising spin models because
they serve  as basic models in condensed matter physics, and focus on
the classical regime to avoid the complication of quantum
entanglement \cite{qgames}.
By treating each site or subset of sites in the Ising-model as a
game-playing agent, we show that
non-cooperative behaviour is indeed possible. In particular, we show the
emergence of a prisoner's dilemma game being `played' within a multi-spin
system.  

We start with some necessary background material. The
two-player prisoner's dilemma game
\cite{games} is characterized by the following payoff matrix:
\[
M:=\begin{array}{|c|c|}  \hline (2,2)&(0,3) \\ \hline (3,0)
& (1,1)
\\
\hline
\end{array} \ .
\] 
The rows designate Rose's choice of strategy (A or B) while the columns
designate Colin's choice of strategy (A or B). The entry $(a,b)$ indicates
that Rose receives payoff $a$ while Colin receives $b$.  A Nash
equilibrium  is the set of strategy choices that provides the optimal
payoff for each player individually: no individual player can improve his
payoff by changing strategy. There is a Nash equilibrium $(1,1)$ with
payoffs 1 to each player. However the global or `Pareto' optimum is
$(2,2)$ wth payoffs 2 to each player. Since the players have no way
of `planning' coordinated moves, the system will end up choosing a payoff
which is macroscopically inferior yet microscopically superior for the
individual players: hence the game is \emph{non-cooperative}. Next we
motivate
the introduction of game-theoretic language to a physical system, by means
of the
following simple example involving two spins with a ferromagnetic spin-spin
interaction:
$H=-2|\uparrow \uparrow\rangle \langle \uparrow \uparrow|+ 2|\uparrow
\downarrow\rangle \langle \uparrow \downarrow |+2 |\downarrow
\uparrow\rangle \langle \downarrow \uparrow | -2|\downarrow
\downarrow\rangle \langle \downarrow \downarrow|$.
At zero temperature and
at equilibrium,  the system will adopt the state with the lowest
energy, i.e. $|\uparrow \uparrow\rangle$ or $|\downarrow \downarrow\rangle$.
(Recall we are ignoring quantum effects such as entanglement for the
present discussion). The energy of the system will now be
$-2$.  In comparison with the other two possible states
$(|\uparrow \downarrow \rangle$ and $|\downarrow \uparrow \rangle)$, the
two particles have gained in stability by losing energy. The negative form
of the Hamiltonian can be used to quantify this gain (and hence loss in
energy) at zero temperature. Hence the system has
gained a payoff of 2 as compared to the value with the spins
separated to infinity.  Because of the system's symmetry, there is no
change if we exchange particles 1 and 2 -- hence we quantify the gain for
each particle by dividing the total gain by two. Hence particles 1 and 2
have gained a payoff of 1. With these definitions,  we may use  the
following game payoff matrix to represent the physical system:
\[
H:=\begin{array}{|c|c|} \hline (1,1) & (-1,-1) \\ \hline (-1,-1) &(1,1) \\
\hline
\end{array}
\] 
which has two equivalent Nash equilibria $(1,1)$.

We now consider the case of two more general subsystems each having
two possible choices of states. These
subsystems may be individual spins, blocks of spins, or more general
subsets of the total many-particle system. For simplicity, we assume the
two subsystems are identical in their composition. We will prove that the
global optimal solutions of the total system will always be at a Nash
equilibrium and hence the interesting dilemma associated with the
prisoner's game cannot unfortunately arise. We suppose that the states of
subsystems 1 and 2 are completely described by the vectors
$\vec{x}$ and $\vec{y}$ respectively. A general Hamiltonian for the
combined  system will have the form: $H_1 +H_2+H_{12}$ where
$H_1$ and $H_2$ are the internal energies of the subsystems when separated
(or equivalently, when the interaction is turned off) while $H_{12}$ is the
interaction term. The symmetry within the system means that we can denote
$H_1=H_2\equiv H$ and $H_{12}=H_{21}\equiv I$.
The condition $H(\vec{x})= H(\vec{y})$ implies that
no one particular state is preferred and so the two subsystems will
have equal incentive to start in
any of the two states before the interaction is turned on.
Since the two
subsystems share the payoff in the interaction term, they each have
the payoff 
$[-H(.)-I(.)/2]$. It follows that whichever
strategy they each adopt, the two subsystems will always have the
same payoffs. To complete the proof, we first define the set of best
replies with respect to $\vec{y}$ for system $1$ to be $B_1(\vec{y}):=\{
\vec{x}^* : P_1(\vec{x}^*, \vec{y}) \geq P_1(\vec{x},\vec{y}), \forall
\vec{x} \}$. In words, given that system 2 has adopted
state-vector $\vec{y}$ then the choice of state-vector
$\vec{x}^*$ by system 1 yields the maximum payoff
$P_1$. The set of best replies is defined similarly for system 2.
Using this notation, a Nash
equilibrium represents an operator profile
$(\vec{x},\vec{y})$ for which
$\vec{x} \in B_1(\vec{y})$ and $\vec{y} \in B_2(\vec{x})$. If
$(\vec{x}^*, \vec{y}^*)$ is a global optimal state, then  by definition
$\vec{x}^* \in B_1(\vec{y}^*)$ and
$\vec{y}^* \in B_2(\vec{x}^*)$ and so
$(\vec{x}^*, \vec{y}^*)$ is a solution at Nash equilibrium.  Therefore, in
this `two-player-two-move' symmetric scenario,  the global optimal solutions
are always at Nash
equilibrium.  We will refer to this situation as {\it cooperative},
noting that this word has a somewhat more general meaning in game theory
\cite{games}. The above analysis indicates that any effective two-body
(i.e. two-subsystem) Hamiltonion system containing just two
strategies/states, is cooperative.

So can macroscopic non-cooperative behavior ever arise? We now show that
it can, if the subsystem under consideration has several possible
configurations and hence more states from which to choose.
In particular, we will search for non-cooperative phenomena in two
subsystems each containing many spins. We concentrate on the specific
example of non-cooperation offered by the prisoner's dilemma, since this is
the only symmetric  two-player two-strategy game with a unique
Nash equilibrium which never coincides with the global optimum~\cite{We95}.
We start by specifying more precisely the criterion for the  existence of
non-cooperative behaviour. A system composed of two-body Ising Hamiltonians
such as
\[ H= a \Big( \ |\uparrow \uparrow\rangle \langle \uparrow \uparrow |-
|\uparrow \downarrow\rangle \langle \uparrow \downarrow |- |\downarrow
\uparrow\rangle \langle \downarrow \uparrow | +|\downarrow
\downarrow\rangle \langle \downarrow \downarrow |\ \Big),
\] may easily be represented as a graph. For example, consider the
following graphical representation of the total Hamiltonian for a
particular five-particle spin system comprising various two-body
interactions:
\[
\begin{array}{crclc}
\uparrow_1 &  \cdots (-2) \cdots & \uparrow_2 \\
\vdots & & \vdots  \\
(3) & & (3)  & & \ \ \ \ \ \ \ \ \ \ \ \ \equiv H_5 \\
\vdots & & \vdots \\
\downarrow_3 & \cdots (-2) \cdots & \downarrow_4 & \cdots (-1)
\cdots  \downarrow_5
\end{array}
\]  where entries in brackets denotes the $a$-parameters in the
five-body Hamiltonian. The total Hamiltonian may then be represented by the
following matrix:
\[
H_5\equiv \left[ \begin{array}{ccccc} 0& -2& 3& 0&0 \\ -2& 0&0&3&0\\
3&0&0&-2&0\\ 0&3&-2&0&-1\\ 0&0&0&-1&0
\end{array} \right] \ .
\]
We now consider the coexistence of two subsystems, each containing
$n$ spins. In order that the total system can be considered as exhibiting
the prisoner's dilemma scenario, the following requirements must be met:
(1)~there are two locally stable states that each subsystem can be in when
they are separated, i.e. in the absence of interaction terms between the
subsystems; (2)~when the two subsystems are put together, i.e. when the
interactions between them are turned on, the final state of the two
subsystems depends on the initial states of these two subsystems; (3)~the
final payoff for each subsystem (as defined above) must reproduce the same
form as the prisoner's dilemma payoff matrix. By symmetry, we require that
the two subsystems are identical and that all interactions are invariant
under the exchange of the two subsystems. Therefore, it must be possible to
write the overall Hamiltonian in the following form:
\begin{equation}
\label{m} M:= \left[ \begin{array}{c|c} X & Y \\ \hline Y & X
\end{array} \right] \ ,
\end{equation} where $[X]$ denotes the Hamiltonian for each system when
isolated, and
$[Y]$ denotes the interaction between the two subsystems. We note that
$X$ and $Y$ must be symmetric matrices in order to render the resulting game
symmetric.  We will represent $\uparrow$ by 1 and $\downarrow$ by $-1$ and
can hence represent the state of the whole system by a string $\delta_1
\ldots
\delta_{2n}$ with $\delta_k= 1$ or $-1$. The total energy of the system
will be
$\frac{1}{2} \sum_{i,j} \delta_i \delta_j M_{ij}$. The payoff
for subsystem 1 will be the negative of the sum corresponding to the upper
half of the matrix $M$, while the payoff for subsystem 2 will be the
negative of the sum corresponding to the lower half of the matrix $M$.
From now on, we will use $A$ and $B$  to denote the two states corresponding
to
the two strategies.

With all the criteria for non-cooperativity laid out, we are now ready to
discuss under what conditions non-cooperative behavior is possible. We will
concentrate on the infinite-range Ising model and will look for the minimal
number of restrictions for two interacting subsystems which each contain $n$
spins. By condition (1), we see that we need to construct two wells with
the same energy at the bottom. In general, this would mean one equality and
$2n$ inequalities. The equality comes from the equal energies for the two
ground
states of the subsystems before interacting, and the $2n$ inequalities come
from the
fact that the whole system is equivalent to a $2n$-dimensional vector space
where
each element has $2n$ neighbors in terms of Hamming distance.
But since we are looking for the minimum number of
restrictions, we may choose the two states corresponding to the bottoms of
the two wells to be adjacent: for example, the state
$|00\dots 00\rangle$ and
$|00\dots 01 \rangle$  (where for simplicity, we are now representing
$\uparrow$ by 0 and
$\downarrow$ by 1). In this case only one equality and $2n-2$ inequalities
need to be satisfied. However the two initial states cannot be adjacent
since we do not want $(A,A)= |00\dots 00\rangle \otimes
|00\dots 00 \rangle$ to be adjacent to $(A,B)= |00\dots 00\rangle
\otimes |00\dots 01 \rangle$; if it were adjacent, then there is an
incentive for the whole system to go from one configuration to another,
thereby
implying that the resulting game may not have well-defined payoffs.
Fortunately,
due to the symmetry in the Ising Hamiltonians, we may replace $00\dots 01$
by
$11\dots 10$. Therefore, we designate
$00\dots 00$ and $11\dots 10$ to be the two initial states (i.e.
strategies)
$A$ and $B$ respectively. We note that since the interactions have infinite
range, it does not matter where the 1s and  0s are located. In terms
of the entries in matrix~(\ref{m}), we have the following equalities and
inequalities:
\begin{eqnarray}
\label{eq}
\sum_{1 \leq j \leq n} x_{nj} &=& 0\\
\label{first}
\sum_{1 \leq j \leq n} x_{kj} &>& 0 \ ,  \ 1\leq k \leq n-1 \\
\sum_{1 \leq j \leq n-1} x_{kj} -x_{kn}&>& 0 \ , \ 1\leq k \leq n-1.
\end{eqnarray}
We now come to condition~2. In order to obtain the least number
of restrictions, we again require that the state $(A,A)$ is at a local
minimum of the whole system, which amounts to $n$ inequalities:
\begin{equation}
\sum_{1 \leq j \leq n} (x_{kj}+y_{kj})> 0 \ ,  \ 1\leq k \leq n.
\end{equation} Similarly, requiring the state $(B,B)$ to be at local
minimum means:
\begin{eqnarray}
\sum_{1 \leq j \leq n-1} (x_{kj} -y_{kj})- x_{kn}-y_{kn}&>& 0 \ , \ 1\leq k
\leq n-1 \\
\label{bb}
\sum_{1 \leq j \leq n-1} (-x_{nj}-y_{nj})- x_{nn}+y_{nn}&>& 0
\end{eqnarray} Using Eq.~\ref{eq}, we can write Eq.~\ref{bb} as:
\begin{equation}
\sum_{1 \leq j \leq n-1} (-y_{nj}) +y_{nn}> 0
\end{equation} If the initial state of the whole system is at state
$(A,B)$, we then require that the first system,
which is at state 
$|00\dots 00\rangle$, transform itself to state $|10 \dots 00\rangle$.
Therefore, the whole system will then be at state $|10 \dots 00\rangle
\otimes |11\dots 10\rangle$. We further require that it is at a local
minimum. Hence the following inequalities apply:
\begin{eqnarray}
\sum_{1 \leq j \leq n-1} (y_{1j}-x_{1j})- y_{1n}&>& 0 \\
\sum_{1 \leq j \leq n-1} (x_{kj} -y_{kj})+ x_{kn}+y_{kn}&>& 0 \ , \ 2\leq k
\leq n \\
\sum_{1 \leq j \leq n-1} (x_{kj} -y_{kj})- x_{kn}-y_{kn}&>& 0 \ , \ 1\leq k
\leq n-1 \\
\sum_{1 \leq j \leq n} y_{nj}&>& 0.
\end{eqnarray} 
Finally, since the prisoner's payoff table is desired, we
need  the following three  inequalities concerning the payoffs in different
scenarios:
\begin{eqnarray}
\label{last}
\sum_{1 \leq j \leq n} y_{nj}-y_{nn}&>& 0\\
\label{final1} -y_{n1}-\sum_{
\stackrel {1\leq k \leq n-1} {2 \leq j \leq n}} y_{kj}&>& 0\\
\label{final2} 2\sum_{1\leq j \leq n-1}x_{1j}+
\sum_{\stackrel{2\leq k \leq n-1} {1 \leq j \leq n-1}}y_{kj}-
\sum_{2 \leq j \leq n-1} y_{nj} &>&0.
\end{eqnarray} Indeed, there are $6n$ inequalities and 1 equality in
total. We know that if there are $6n+1$ arbitrary parameters, then the
above system of inequalities has a solution \cite{Fa}. However by exploiting
many of the redundancies inherent in the system, we will now show that only
$n+2$ non-zero parameters are actually needed. Table 1 summarizes the
ranges of these inequalities.
From Table~1, we see that for $k=1$ we may set
every parameter to zero {\em except} $b_1,d_1$ and $f_1$, which satisfy the
following inequalities:
\begin{eqnarray} b_1 &>& 0 \\ d_1 &>& 0 \\ f_1 & <&0 \\ d_1 &>& b_1+f_1.
\end{eqnarray} For $k=n$,  since we have set $f_1$ to be non-zero and $Y$
is symmetric, we have $d_n=f_1$. Hence we may now set every parameter to
zero except $d_n, e_n$ and $f_n$.  The following inequalities then follow:
\begin{eqnarray} e_n &>& -f_1 \\ f_n & >&0 \\ f_n &>& e_n+f_1.
\end{eqnarray} For $2\leq k \leq n-1$, we may set all parameters to zero
except $b_k$. Recall that we have already set $e_n>0 $ hence we may now also
set 
$b_k>e_n$. Finally, we come back to Eqs~\ref{final1} and \ref{final2},
which yields two further inequalities:
\begin{eqnarray} -2f_1&>&e_n \\ 2b_1 &>& f_1.
\end{eqnarray}
 
The following form corresponds to two $n=3$ spin subsystems, and is
one of many possible scenarios which satisfies the above requirements:
\[ X:=
\left[
\begin{array}{ccc} 0& 2& 0 \\ 2 & 0&0  \\ 0& 0 &0
\end{array}
\right] \ ; \ \ \ \ Y:=
\left[
\begin{array}{ccc} 2&0&-1 \\ 0& 0&1.5 \\ -1&1.5 &2 \\
\end{array}
\right]\ ,
\] 
The corresponding payoff matrix is:
\begin{equation}
\label{prisoner}
M:=\begin{array}{|c|c|} \hline (4.5,4.5)&(1,5) \\ \hline (5,1)&(3.5,3.5)\\
\hline
\end{array} \ .
\end{equation}
By considering the number of non-zero parameters required, one can
deduce that an $n=3$-spin subsystem `playing' against another $n=3$-spin
subsystem is the {\em smallest} total system which exhibits
non-cooperative behaviour within the Ising framework: the phrase `two's
company but three's a crowd' therefore comes to mind \cite{michael}. 
The above discussion
has been mostly concerned with infinite-range interactions, however we note
that one may also analyse the positions of the  non-zero parameters in
order to discuss various other models such as nearest-neighbor interaction
models, etc. 

We now return to discuss whether
the Ising-based games considered here exhibit similar behavior to that
observed in spatial games. One interesting phenomenon observed in the
spatial prisoner's game is the coexistence of different strategies up to
some critical point~\cite{NM92}. It turns out that this same phenomenon
also occurs for our Ising games. To show this, we
consider a large system with many
$n$-spin subsystems such that each subsystem is {\it playing a prisoner's
game with its immediate neighbors}. In order that the subsystems evolve
their strategies, we add in fluctuations which we  assume to be
thermal. At low enough temperature, only  very
few subsystems (which are statistically far apart) will have the
chance to change their states at a given timestep. These subsystems
will, with high probability, change to a state which minimizes their
free energy -- we assume this to be close to the energy that we use to
construct the payoff matrix table.  Regardless of what their neighbors'
states actually are, these subsystems will always choose state $B$. (If
their neighbors are in state $A$, they can exploit their neighbors by
adopting state
$B$, whereas if their neighbors are in state
$B$, then adopting state $B$ is the best they can do).  We note that this is
in marked contrast to previous assumptions used in spatial games \cite{NM92,
ST98}, where agents were usually assumed to imitate the
strategy of the best player in their neighborhood; since a
subsystem cannot `know' the surrounding subsystems' payoffs, it can only
maximize its own payoff given any particular situation.  At low
enough temperature, the whole system will evolve to the situation where
every subsystem is in state
$B$ given any initial  condition. This resembles the phenomenon of
condensation. By contrast, if the temperature is high enough that
many subsystems are allowed to change their states simultaneously but still
low enough that the free energies exhibit a payoff matrix similar to the
prisoner's dilemma (see Eq. (\ref{prisoner})), then with high
probability neighboring sites or blocks of neighboring sites may be able
to change their states together. Then with high probability, the subsystems
in the block will all end up in state
$A$ because it maximizes the payoff for the whole block. Therefore at high
temperature, the majority of sites will adopt state $A$. Just as for the
spatial games considered previously in the literature, a whole
spectrum of scenarios is possible for the Ising games even though the
detailed dynamics of these two models may  have different origins.
Beside thermal fluctuations, we may also consider a fluctuation in the
payoffs. In Ref.~\cite{NM92}, Nowak and May used the following payoff
matrix:
\begin{equation}
M:=\begin{array}{|c|c|} \hline (1,1)&(0,b) \\ \hline (b,0)&(0,0)\\ \hline
\end{array} \
\end{equation} where $b>1$. They found that a region where the
coexistence of both strategies is encouraged, appears for $1.8<b<2$.  In
our Ising games, this is also what one expects to happen: when $b$
is close to the prescribed range, the  configurations
$(A,A), (A,B)$ and $(B,A)$ will all have similar global energies, hence all
three configurations are preferred at high enough temperature such that
neighboring subsystems can evolve together.  This similarity in the
region of interest further strengthens the tie between spatial games and
condensed matter systems.

Finally we note that the kind of fluctuation considered here
is different from the conventional models considered
in the fluctuating systems literature associated
with spin glasses and other condensed matter systems. Instead it
is usually assumed that given enough time, a system will achieve
equilibrium: the fluctuation of the whole system is
taken to mean that the subsystems fluctuate as well, hence the whole
system has an incentive to move to equilibrium. Here we encounter
something different: given the initial conditions, some subsystems can
actually benefit from  the fluctuation and so have no incentive to increase
the global payoff. For more general models, such as the Heisenberg model,
one would  expect quantum fluctuations to play an
essential role, thereby suggesting a possible future role for quantum game
theory
\cite{qgames}.

\newpage

\begin{table}
\label{tab}
\caption{Table of inequalities Eq. (\ref{first})--(\ref{last}) where
$a_kx_{k1}+b_k
\sum_{2 \leq j \leq n-1} x_{kj}+c_k x_{kn}+d_k y_{k1}+e_k \sum_{2 \leq j
\leq n-1} y_{kj}+f_k y_{kn}>0$.}
\vskip0.5in
\begin{tabular}{|c|c|c|c|c|c|c|} \hline
&$a_k$&$b_k$&$c_k$&$d_k$&$e_k$&$f_k$ \\
\hline
$ k=1$&1&1&1&1&1&1       \\  &1&1&--1&--1&--1&--1 \\  &1&1&1&0&0&0       \\
&1&1&--1&0&0&0      \\ &--1&--1&0&1&1&--1 \\
\hline
$k=n$ &0&0&0&--1&--1&1 \\ &0&0&0&1&1&1 \\ &0&0&0& 1&1&0\\
 \hline
$2\leq k \leq n-1$& 1&1&1&1&1&1 \\ &1&1&--1&--1&--1&--1 \\  &1&1&1&0&0&0 \\
&1&1&--1&0&0&0\\ &1&1&1&--1&--1&1\\ \hline
\end{tabular}
\end{table}

\end{document}